\documentclass[12pt]{iopart}
\usepackage{iopams}
\usepackage{graphics}
\pdfoutput=1
\usepackage[latin1]{inputenc}
\usepackage{amscd}
\usepackage{bbm}
\usepackage[all,cmtip]{xy}
\usepackage{color}

\newcommand{\mathd}{\mathrm{d}}
\newcommand{\mathe}{\mathrm{e}}

\newcommand{\tmmathbf}[1]{\ensuremath{\boldsymbol{#1}}}
\newcommand{\tmop}[1]{\ensuremath{{#1}}}
\newcommand{\tmtextbf}[1]{{\bfseries{#1}}}
\newcommand{\tmtextit}[1]{{\itshape{#1}}}

\begin{document}

\title[]{A classical appraisal of quantum definitions of non-Markovian
dynamics}

\author{Bassano Vacchini \dag \ddag}

\address{\dag Dipartimento di Fisica, Universit{\`a} degli Studi di
Milano, Via Celoria 16, I-20133 Milan, Italy\\
\ddag
INFN, Sezione di Milano, Via Celoria 16, I-20133
Milan, Italy}



\ead{bassano.vacchini@mi.infn.it}

\begin{abstract}
We consider the issue of non-Markovianity of a quantum dynamics starting from
a comparison with the classical definition of Markovian process. We point to
the fact that two sufficient but not necessary signatures of non-Markovianity
of a classical process find their natural quantum counterpart in recently
introduced measures of quantum non-Markovianity. This behavior is analyzed in
detail for quantum dynamics which can be built taking as input a class of
classical processes.
\end{abstract}


\section{Introduction\label{sec:intro}}

In the field of open quantum systems the interaction with an external
environment introduces a stochasticity element in the dynamics, which has
typically been described as a quantum process in analogy with classical
processes {\cite{Breuer2002}}. Quantum Markovian processes have been
understood as described by a master equation in Lindblad form
{\cite{Gorini1976a,Lindblad1976a}}, which grants complete positivity.
Furthermore these dynamics can be understood as an average over trajectories
corresponding to suitable measurements continuous in time on the side of the
environment, thus providing a direct link with the description of quantum
measurement {\cite{Barchielli2009}}.

However the very notion of non-Markovianity for a quantum dynamics has long
remained vague, being usually associated with the occurrence of revivals or
non exponential relaxation, apart from mathematical work in which a proper
definition of a quantum process has been addressed
{\cite{Lindblad1979a,Accardi1982a}}. Most recently, different approaches have
been devised in order to assess and quantify non-Markovianity in the dynamics
of an open quantum system, looking at states rather than at multitime
correlation functions. One approach, based on the idea that memory can be seen
as a backflow of information from the environment to the system, studies the
distinguishability among states in the course of the dynamics, identifying
non-Markovianity with revivals in the distinguishability. The amount of
non-Markovianity then depends on frequency and relevance of these revivals
{\cite{Breuer2009b}}. Note that this approach is actually valid in order to
estimate non-Markovianity of the dynamics in any statistical theory, simply
considering the mathematical representation of the set of states and a
distance on it, which should be a contraction under the action of positive
probability preserving maps. Another approach relies on the divisibility over
arbitrary intermediate time intervals of the evolution map into well defined
maps preserving positivity and probability. It estimates non-Markovianity
quantifying the failure of this property {\cite{Rivas2010a}}.

In recent work it has been pointed out that these two approaches can be
connected to certain signatures of non-Markovianity to be detected at the
level of the one-point probability in a stochastic process
{\cite{Vacchini2011a}}, and a few examples of this behavior have been
considered building on a class of stochastic processes known as semi-Markov
processes {\cite{Breuer2008a,Breuer2009a}}. In this contribution we will
consider the class of maps which can be built within this framework by
considering quantum dynamics determined by a generic unital stochastic map and
an arbitrary waiting time distribution. We will exploit these examples to
study in detail the different performance of the two measures and the variety
of possible behavior which can be obtained. This class of dynamics will also
allow us to point to the proper relationship between divisibility of the time
evolution in terms of completely positive maps and behavior of the
coefficients in the time-convolutionless form of the equations of motion.

The paper is organized as follows. In Sect.~\ref{sec:mar} we introduce the
definition of classical Markov process, together with previous proposals to
extend this definition to the quantum realm. We further consider two
sufficient signatures of non-Markovianity to be read from the one-point
probability density. In Sect.~\ref{sec:qnmd} it is shown how the extension of
these signatures to the quantum case recovers two recent proposals for the
definition of a non-Markovian dynamics, which are studied and compared for a
general class of examples. Conclusions and final remarks are presented in
Sect.~\ref{sec:ceo}.

\section{Markovian and non-Markovian processes\label{sec:mar}}

The notion of stochastic process is of the greatest importance in the
description of phenomena in which randomicity appears. Due to interaction with
the environment open quantum systems do provide a natural setting in which a
stochastic description appears. This in turns leads to the difficult question
about how to properly characterize a quantum stochastic process, addressed in
different ways by physicists {\cite{Gardiner1991,Breuer2002}} and
mathematicians {\cite{Lindblad1979a,Accardi1982a}}. Within this framework a
further important question is how to define Markovianity of a process and how
to associate a memory to the various phenomena.

In the classical case one has a precise definition of Markovian process,
relying on the knowledge of all finite dimensional distributions of the
process, which is best formulated in terms of conditional probabilities
{\cite{Gillespie1998a}}
\begin{eqnarray}
  p_{1| n} (x_n, t_n |x_{n - 1}, t_{n - 1} ; \ldots ; x_0, t_0) & = & p_{1|1}
  (x_n, t_n |x_{n - 1}, t_{n - 1})  \label{eq:mp}
\end{eqnarray}
with $t_n \geqslant t_{n - 1} \geqslant \ldots \geqslant t_1 \geqslant t_0$.
It tells that the probability for the random variable to assume the value
$x_n$ at time $t_n$ only depends on the last assumed value, and not on
previous ones, thus properly formalizing the notion of lack of memory. The set
of conditions Eq.~(\ref{eq:mp}) lead in particular to the Chapman-Kolmogorov
equation obeyed by the two point conditional probability, also called
propagator
\begin{eqnarray}
  p_{1|1} (x, t|y, s) & = & \sum_z p_{1|1} (x, t|z, \tau) p_{1|1} (z, \tau |y,
  s),  \label{eq:ck}
\end{eqnarray}
valid for $t \geqslant \tau \geqslant s$. A solution of this equation fully
characterizes a Markovian process in that all finite dimensional distributions
can be obtained given the initial probability distribution. Since the
propagator is on its turn determined from the dynamics of all mean values,
this means that the whole process is known from the dynamics of mean values, a
result known as regression theorem {\cite{Lax1968a}}. All multitime
correlation functions of the process are thus fixed from the propagator and
therefore from the mean values. Indeed Lindblad introduced a definition of
quantum Markovian process by relying on the validity of the quantum regression
theorem {\cite{Lindblad1979a}}. Consider a system $S$ interacting with an
environment $E$ according to a unitary evolution $U (t)$ and multitime
correlation functions of the form
\begin{eqnarray}
  \hspace{-2cm} \langle \mathcal{M}_n (t_n) \ldots \mathcal{M}_1 (t_1) \rangle
  & = & \tmop{Tr}_S \tmop{Tr}_E \{\mathcal{M}_n U (t_n - t_{n -
  1})\mathcal{M}_{n - 1} \ldots \mathcal{M}_1 U (t_1) \rho_S \otimes \rho_E
  \}, \nonumber
\end{eqnarray}
where $\mathcal{M}_i$ denotes a quantum operation corresponding to measurement
of a certain system observable, e.g. $\mathcal{M}_{} (\varrho) = \sum_k P_k
\varrho P_k$ for the von Neumann measurement of the self-adjoint operator $A =
\sum_k a_k P_k$. If the quantum regression formula applies such multitime
correlation functions can be expressed as
\begin{eqnarray}
  \langle \mathcal{M}_n (t_n) \ldots \mathcal{M}_1 (t_1) \rangle & = &
  \tmop{Tr}_S \{\mathcal{M}_n \Phi (t_n - t_{n - 1})\mathcal{M}_{n - 1} \ldots
  \mathcal{M}_1 \Phi (t_1) \rho_S \}, \nonumber
\end{eqnarray}
with $\Phi (t)$ completely positive maps acting on the system only. In the
weak and singular coupling limit these maps can be shown to satisfying a
semigroup composition law {\cite{Dumcke1983a}}, so that they can be expressed
with a generator in Lindblad form. As a result in particular mean values and
correlations do obey the same dynamical equations. This result, though by no
means obvious for the behavior of the reduced dynamics of an open system, is a
basic working tool in open quantum system theory {\cite{Breuer2002}}, and the
extension of the validity of the regression formula to the non-Markovian
setting is an issue of great relevance. The problem has been recently
addressed for a special class of non-Markovian evolutions known as generalized
Lindblad structure {\cite{Budini2004a,Breuer2007a,Vacchini2008a}}, showing
however that regression formula are generally not valid {\cite{Budini2005b}}.

\subsection{Witnesses and quantifiers of non-Markovianity\label{sec:measure}}

The validity of the Chapman-Kolmogorov equation for a Markovian process
entails two simple consequences on the behavior of the propagator, which are
at the heart of two recently introduced quantum measures of non-Markovianity
{\cite{Breuer2009b,Rivas2010a}}. Let us consider for the sake of simplicity a
random variable taking values on a finite set, so that the propagator from
time $s$ to time $t$ can be written as a stochastic matrix $\Lambda (t, s)$
and the state of the system at a given time $t$ can be expressed as a
probability vector $\tmmathbf{p}(t)$. As distance among probability vectors we
consider the Kolmogorov distance {\cite{Fuchs1999a}}
\begin{eqnarray}
  D_K (\tmmathbf{p}^1 (t), \tmmathbf{p}^2 (t)) & = & \frac{1}{2} \sum_i |p_i^1
  (t) - p_i^2 (t) |. \nonumber
\end{eqnarray}
Upon validity of the Chapman-Kolmogorov equation one has $\tmmathbf{p}(t) =
\Lambda (t, s)\tmmathbf{p}(s)$ for any $0 \leqslant s \leqslant t$, and
exploiting the definition of stochastic matrix this entails that the
Kolmogorov distance among two probability distributions following a Markovian
dynamics do decrease monotonously in time
\begin{eqnarray}
  D_K (\tmmathbf{p}^1 (t), \tmmathbf{p}^2 (t)) & \leqslant & D_K
  (\tmmathbf{p}^1 (s), \tmmathbf{p}^2 (s)) \hspace{2em} \forall t \geqslant s.
  \nonumber
\end{eqnarray}
This relation naturally provides a necessary condition for the Markovianity of
a stochastic process, stating that in a Markovian process the one-time
probability densities or distribution functions evolving from two distinct
initial situations do get less and less distinguishable. A non monotonic
behavior in time of the Kolmogorov distance among two states thus provides a
witness of the non-Markovianity of the process. Of course this condition now
appears as a sufficient condition only, or if one prefers a different
definition of Markovianity.

Let us stress that the Kolmogorov distance actually corresponds to the $L_1$
distance among probability vectors, thus being a natural distance in any
statistical theory. Indeed a statistical theory relies on the existence of two
spaces, one dual to the other, in which states and observables do live. The
specific choice of spaces and their commutativity or non commutativity do fix
the statistical structure of the theory
{\cite{Holevo1982,Ludwig1983,Vacchini2007a,Vacchini2010a}}. In the case of
classical mechanics one considers the $L_1$ space of probability distributions
on phase space, while the observables are given by the $L_{\infty}$ space of
bounded functions. Considering quantum mechanics the dual Banach spaces are
given by the space of trace-class operators with the trace norm topology, in
which states are described by statistical operators, and the space of bounded
operators with the uniform norm, in which to consider the observables of the
theory. The Kolmogorov distance thus naturally becomes the trace distance
among statistical operators.

The validity of the Chapman-Kolmogorov equation for a Markov process entails
another important consequence at the level of the one-point probability
density, namely rewriting Eq.~(\ref{eq:ck}) in terms of stochastic matrices
for a finite dimensional process one has $\Lambda (t, s) = \Lambda (t, \tau)
\Lambda (\tau, s)$ for any $s \leqslant \tau \leqslant t$. Taking $s$ as an
initial time set equal to zero one has the relation
\begin{eqnarray}
  \Lambda (t, 0) & = & \Lambda (t, \tau) \Lambda (\tau, 0)  \label{eq:cpdiv}
\end{eqnarray}
where the crucial fact is that each $\Lambda$ is a well-defined stochastic
matrix sending any probability vector to a probability vector. For a Markov
process these matrices are fixed by the transition probabilities.
Eq.~(\ref{eq:cpdiv}) then describes a divisibility property which provides a
witness of Markovianity. However the violation of this property only provides
a necessary but not sufficient condition in order to ascertain Markovianity.
Indeed also for non-Markovian processes one might find a collection of
stochastic matrices obeying Eq.~(\ref{eq:cpdiv}), which nevertheless do not
coincide with the transition probabilities of the process
{\cite{Hanggi1977a}}. Otherwise stated these collection of stochastic matrices
can be taken as the transition probabilities of a Markov process which at the
level of the one-point probabilities cannot be distinguished from the original
one.

These signatures of non-Markovianity for a classical process correspond to
criteria used to define a non-Markovian dynamics in the quantum case in
{\cite{Breuer2009b}} and {\cite{Rivas2010a}}, by considering
distinguishability as quantified by the trace distance, and divisibility as
characterized by a composition in terms of completely positive maps
respectively. In order to further assign a weight to the deviation from the
Markovian behavior, relying on these signatures two measures of
non-Markovianity have been introduced, which essentially assign a weight to
the time intervals in which either trace distance grows or divisibility, at
the quantum level to be understood as divisibility in terms of completely
positive maps, fails. These weights can actually be assigned in different
ways. For example one can consider quantifiers of distinguishability on the
state space other than trace distance, provided they are contractions with
respect to the action of positive maps {\cite{Laine2010a,Dajka2011a}}, e.g.
relative entropy which appears in a natural way in certain dissipative systems
{\cite{Vacchini2004a,Vacchini2009a}}.

\subsection{Semi-Markov processes\label{sec:smp}}

Let us now consider a class of stochastic processes for which a simple
characterization is available, and which include both Markovian and
non-Markovian processes, namely semi-Markov processes {\cite{Feller1971}}.
Such processes arise considering transitions among different sites determined
from certain jump probabilities, fixed by a stochastic matrix as in a
Markovian chain, and random waiting times between jumps determined by site
dependent waiting time distributions. They are characterized via a semi-Markov
matrix
\begin{eqnarray}
  (Q)_{mn} (\tau) & = & (\Pi)_{mn} f_n (\tau),  \label{eq:smm}
\end{eqnarray}
whose entries give the probability density for a jump from site $n$ to site
$m$ in a given time $\tau$. The transition probabilities $\pi_{mn}$ build up a
stochastic matrix, while the $f_n (\tau)$ are the waiting time distributions
which actually characterize whether the process is Markovian or not according
to the classical definition. Indeed such a process is Markovian only if all
waiting time distributions are memoryless, that is exponentially distributed
\begin{eqnarray}
  f_n (\tau) & = & \lambda_n \mathe^{- \lambda_n \tau} .  \label{eq:expwtd}
\end{eqnarray}
In all other cases such processes are non-Markovian. This simple
characterization of semi-Markov processes, together with the fact that their
transition probabilities can be obtained as solution of a generalized master
equation with a memory kernel, allows to exemplify the meaning of the
signatures of non-Markovianity introduced above. In the case in which all
waiting time distributions are equal, though otherwise arbitrary, the
transition probability $T (t, 0)$ of such a process obeys a closed
integrodifferential equation given by
\begin{eqnarray}
  \frac{\mathd}{\tmop{dt}} T (t, 0) & = & \int_0^t \mathd \tau (\Pi -
  \mathbbm{1}) k (\tau) T (t - \tau, 0),  \label{eq:fk}
\end{eqnarray}
where $k (\tau)$ is a memory kernel fixed by $f_{} (\tau)$ according to the
relationship
\begin{eqnarray}
  f (\tau) = \int_0^{\tau} \mathd tk (\tau - t) g (t), & \hspace{1em} & g (t)
  = 1 - \int_0^t \mathd \tau f (\tau),  \label{eq:k-sur}
\end{eqnarray}
where $g (t)$ is the survival probability associated to $f (\tau)$.

We now consider a two-dimensional system and take two equal waiting time
distributions, so that the semi-Markov matrix of Eq.~(\ref{eq:smm}) can be
written
\begin{eqnarray}
  Q (\tau) & = & \left(\begin{array}{cc}
    \pi & \sigma\\
    1 - \pi & 1 - \sigma
  \end{array}\right) f (\tau),  \label{eq:qbisto}
\end{eqnarray}
with $\pi$, $\sigma$ the jump probabilities from one site to the other.
Starting from Eq.~(\ref{eq:fk}) for the case $\pi = \sigma = 1 / 2$, one can
express the transition probability as a function of the survival probability
only, and using $T (t, s) = T (t, 0) T^{- 1} (s, 0)$ one obtains the matrices
\begin{eqnarray}
  T (t, s) & = & \frac{1}{2} \left(\begin{array}{cc}
    1 + g (t) / g (s) & 1 - g (t) / g (s)\\
    1 - g (t) / g (s) & 1 + g (t) / g (s)
  \end{array}\right),  \label{eq:12ts}
\end{eqnarray}
which connect probability vectors at a time $s$ with probability vectors at a
later time $t$. Since $g$ is a survival probability, these matrices are
stochastic matrices for any couple $t \geqslant s$, independently on the fact
that the associated semi-Markov process is Markovian only if $f$ is
exponentially distributed. Thus for any choice of $f (\tau)$ different from
the exponential one has an example of process which is non-Markovian in the
classical sense, but still whose one-point probabilities do contract in a
monotonous way with respect to the Kolmogorov distance and are connected by
stochastic matrices. This indeed shows that contractivity under the Kolmogorov
distance and divisibility are sufficient but not necessary criteria in order
to detect non-Markovianity in the classical sense. \

\begin{figure}[tb]
  \resizebox{!}{60mm}{\includegraphics{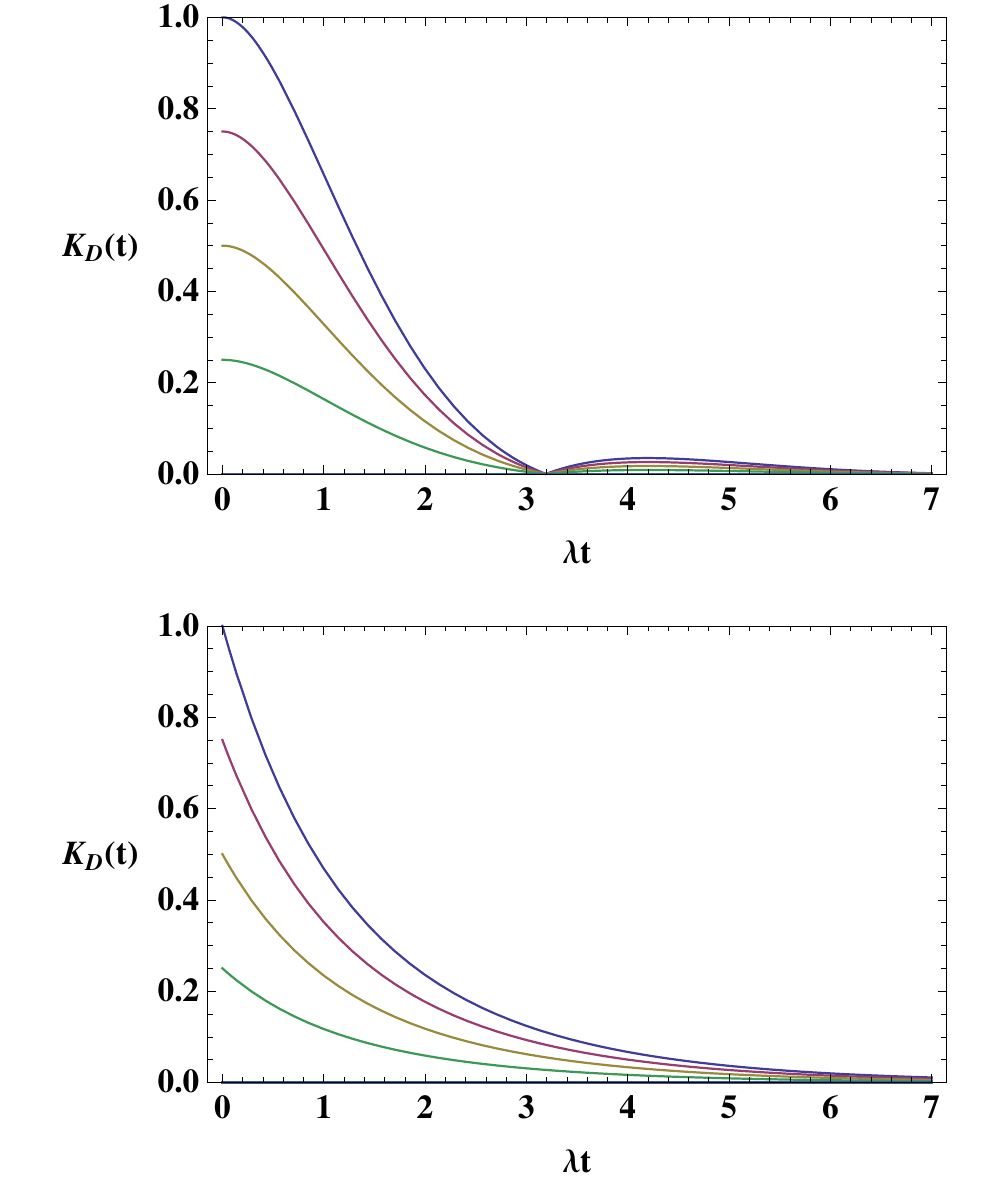}}{\hspace{1em}}\resizebox{60mm}{!}{\includegraphics{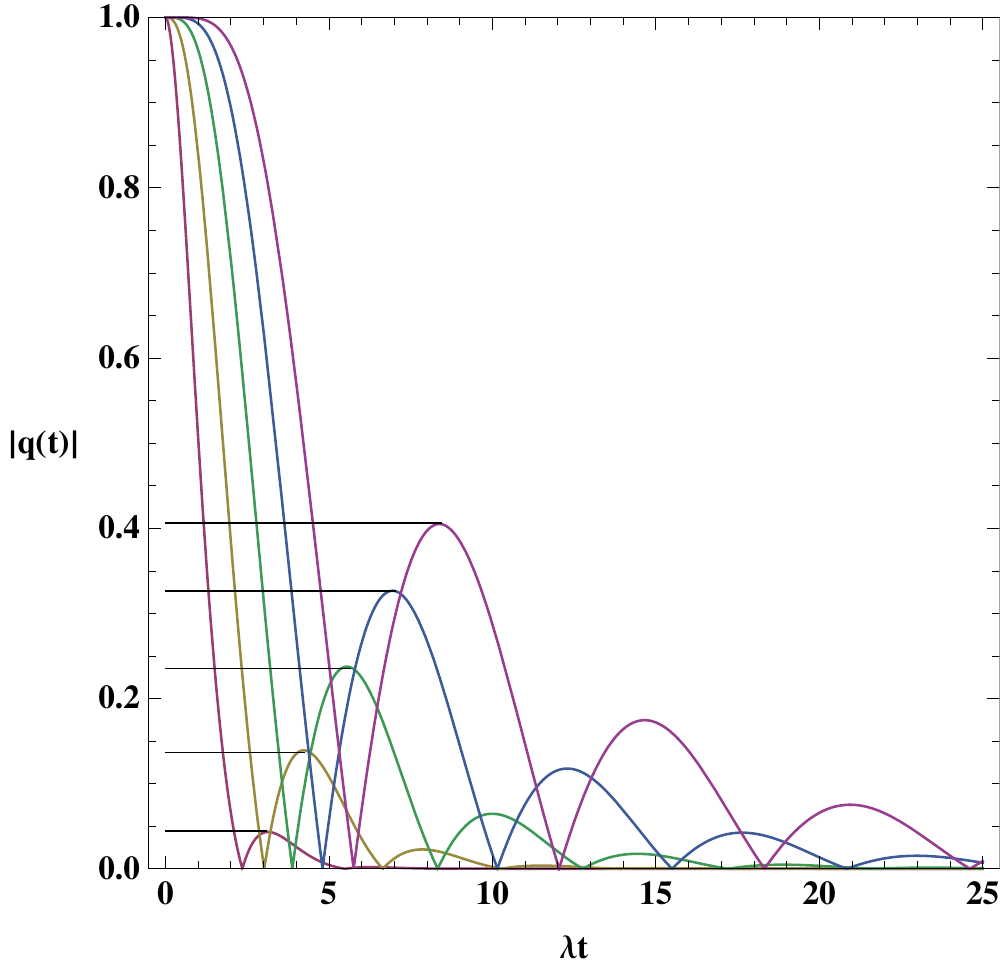}}
  \caption{\label{fig:kolmogorov-qm}(Color online) (top left) Kolmogorov
  distance for a classical semi-Markov process characterized by $\pi = 0$ and
  $\sigma = 1$, and a waiting time distribution given by the convolution of
  two exponential ones with $\lambda_2 = 0.5 \lambda_1$. The different
  trajectories correspond to initial probability vectors such that $|p_1^1 (0)
  - p_1^2 (0) |$ are equally distributed between 0 and 1. Revivals clearly
  appear. (bottom left) Kolmogorov distance for a semi-Markov process with the
  same waiting time distribution but $\pi = \sigma = 1 / 2$. The distance
  decreases monotonically. In both cases the process is non Markovian
  according to the classical definition. (right) The modulus of the functions
  $q_m$, corresponding to the convolution of $m$ equal exponential
  distributions, for $m$ up to 6 and in the case of a semi-Markov process with
  $\pi = 0$ \ and $\sigma = 1$. The horizontal lines denote the height of the
  first non trivial maximum of $|q_m \left( t \right) |$, quantifying the
  non-Markovianity of the first interval in which trace distance grows.
  
  }
\end{figure}

As a complementary situation, let us consider the case $\pi = 0$, $\sigma =
1$, so that once in a state the system jumps with certainty to the other, thus
obtaining
\begin{eqnarray}
  T (t, s) & = & \frac{1}{2} \left(\begin{array}{cc}
    1 + q (t) / q (s) & 1 - q (t) / q (s)\\
    1 - q (t) / q (s) & 1 + q (t) / q (s)
  \end{array}\right) .  \label{eq:1ts}
\end{eqnarray}
The role of the survival probability $g (t)$ is here replaced by the quantity
$q (t)$ whose Laplace transform reads
\begin{eqnarray}
  \hat{q} (u) & = & \frac{1}{u}  \frac{1 - \hat{f} (u)}{1 + \hat{f} (u)} .
  \nonumber
\end{eqnarray}
Recalling that the probability for $n$ jumps in a time $t$ for a waiting time
distribution $f (t)$ is given by
\begin{eqnarray}
  p_n (t) = \int^t_0 \mathd \tau f (t - \tau) p_{n - 1} (\tau), &  & 
  \label{eq:pn}
\end{eqnarray}
so that $\hat{p}_n (u) = \hat{p}_0 (u) \hat{f}^n (u)$, one finally has
\begin{eqnarray}
  q (t) & = & \sum^{\infty}_{n = 0} p_{2 n} (t) - \sum^{\infty}_{n = 0} p_{2 n
  + 1} (t) .  \label{eq:qdiff}
\end{eqnarray}
The quantity $q (t)$ therefore expresses the difference between the
probability to have an even or an odd number of jumps. It immediately appears
that this quantity is not necessarily positive, so that the matrices defined
by Eq.~(\ref{eq:1ts}) are not necessarily stochastic matrices, and apart from
the case of an exponential waiting time distribution, corresponding to a truly
Markovian process, different expression of $f (\tau)$ may or may not lead to \
contractivity under the Kolmogorov distance and divisibility. These possible
behaviors are considered in Fig.~\ref{fig:kolmogorov-qm}, where the Kolmogorov distance for two non-Markovian classical
process is considered.

\section{Quantum non-Markovian dynamics\label{sec:qnmd}}

The characterization of well-defined classes of quantum time evolutions which
have the property of being completely positive, though not in Lindblad form,
is a highly non trivial task. A whole class of completely positive dynamics
can be obtained considering a quantum generalization of the classical
semi-Markov processes introduced in Sect.~\ref{sec:smp}, also allowing for the
connection and comparison with a classical process. Such dynamics are given by
the solution of integrodifferential equations with a memory kernel which is
formally of Lindblad type {\cite{Breuer2008a,Breuer2009a}}. For the case in
which the dynamics allows for a clearcut probabilistic reading such equations
can be written in the form {\cite{Budini2004a}}
\begin{eqnarray}
  \frac{\mathd}{\tmop{dt}} \rho (t) & = & \int^t_0 d \tau k (t - \tau) [
  \mathcal{E} - \mathbbm{1}] \rho (\tau) .  \label{eq:ms}
\end{eqnarray}
Indeed if the function $k (t)$ can be associated to a well defined waiting
time distribution $f (t)$ according to the relation Eq.~(\ref{eq:k-sur}), then
the time evolution of the solution $\rho (t)$ of Eq.~(\ref{eq:ms}) can be
expressed as repeated actions of the completely positive and trace preserving
map $\mathcal{E}$, also called stochastic map, distributed in time according
to the renewal process fixed by $f (t)$
\begin{eqnarray}
  \Lambda (t, 0) \rho (0) & = & \sum^{\infty}_{n = 0} p_n (t) \mathcal{E}^n
  \rho (0),  \label{eq:gspl}
\end{eqnarray}
where $p_n (t)$ is defined in Eq.~(\ref{eq:pn}). Note that Markovianity or
non-Markovianity of this quantum time evolution, according to either criterion
of Sect.~\ref{sec:measure}, at variance with the classical definition depend
on both elements of the couple $\{ \mathcal{E}, f (t)\}$. More precisely,
while the only truly non-Markovian waiting time distribution corresponding to
the exponential leads to a delta correlated kernel in Eq.~(\ref{eq:ms}), and
therefore to a Lindblad equation, a generic waiting time distribution might
still lead to a Markovian dynamics depending on the completely positive trace
preserving map $\mathcal{E}$ as we shall see below.

\subsection{Unital stochastic maps}

Let us first leave $f \left( t \right)$ unspecified and consider as
$\mathcal{E}$ the socalled Pauli channel
\begin{eqnarray}
  \mathcal{E} \left[ \rho \right] & = & \lambda_0 \rho + \lambda_x \sigma_x
  \rho \sigma_x + \lambda_y \sigma_y \rho \sigma_y + \lambda_z \sigma_z \rho
  \sigma_z, \nonumber
\end{eqnarray}
with $\tmmathbf{\lambda} \equiv \left( \lambda_0, \lambda_x, \lambda_y,
\lambda_z \right)$ a probability distribution, $\lambda_i \geqslant 0$ and
$\sum_i \lambda_i = 1$. As discussed in {\cite{King2001a}} up to unitary
transformations this expression provides the most general unital stochastic
map, which preserves both trace and identity for $\mathcal{H =
\mathbbm{C}}^2$. The map $\Lambda (t, 0)$ associated to $\mathcal{E}$ through
Eq.~(\ref{eq:gspl}) can be expressed in terms of a basis of superoperators
acting on operators in $\mathbbm{C}^2$. The standard representation is
\begin{eqnarray}
  \Lambda [\rho] & = & \sum^3_{k, l = 0} \Lambda_{kl} \tau_k \tmop{Tr}_S
  [\tau_l^{\dag} \rho],  \label{eq:mape}
\end{eqnarray}
with $\{\tau_k \}_{k = 0, 1, 2, 3}$ a basis of operators on $\mathbbm{C}^2$
and $\Lambda_{kl} = \tmop{Tr}_S \{\tau_k^{\dag} \Lambda [\tau_l]\}.$ The most
convenient choice of orthonormal basis corresponds to $\tau_i =
\frac{1}{\sqrt{2}} \sigma_i$, with $\sigma_0 = \mathbbm{\mathbbm{1}}$ and
$\sigma_i$ the standard Pauli matrices. These basis elements are eigenoperator
of the Pauli channel. The eigenvalues are given by $\tmmathbf{\mu}=
A\tmmathbf{\lambda}$, with $A_{0 i} = A_{i 0} = 1$ and $A_{jk} = 2 \delta_{jk}
- 1$ for $j, k = 1, 2, 3$, so that $\mu_0 = 1$ and $- 1 \leqslant \mu_i
\leqslant 1$. In order to determine the map $\Lambda \left( t, 0 \right)$ we
consider its action on the basis elements, given by
\begin{eqnarray}
  \Lambda \left( t, 0 \right) \left[ \sigma_i \right] & = & \lambda_i \left( t
  \right) \sigma_i,  \label{eq:elli}
\end{eqnarray}
where the quantities $\lambda_i \left( t \right)$ are defined as
\begin{eqnarray}
  \lambda_i \left( t \right) & = & \sum^{\infty}_{n = 0} p_n (t) \mu^n_i, 
  \label{eq:lt}
\end{eqnarray}
and thus correspond to the generating function of the discrete probability
distribution $p_n (t)$ evaluated at $\mu_i$, in particular $\lambda_0 \left( t
\right) = 1$, while $- 1 \leqslant \lambda_i \left( t \right) \leqslant 1$.
Note that according to Eq.~(\ref{eq:elli}) the standard representation of
statistical operators on the Bloch sphere the action of the map $\Lambda
\left( t, 0 \right)$ transforms the surface of the sphere into ellipsoids
whose axes have a time dependent length $\left| \lambda_i \left( t \right)
\right|$. In the classical setting Markovianity of the semi-Markov process is
only obtained for an exponential waiting time distribution, so that the $p_n
(t)$ are a Poisson distribution. In this case the generating function reads
\begin{eqnarray}
  \lambda_i \left( t \right) & = & \mathe^{- \left( 1 - \mu_i \right) \lambda
  t},  \label{eq:gfp}
\end{eqnarray}
where $\lambda$ is the parameter of the exponential distribution, thus
corresponding to ellipsoids whose axes shrink monotonously according to an
exponential law. Exploiting Eq.~(\ref{eq:pn}) the Laplace transform of the
$\lambda_i \left( t \right)$ can be directly expressed in terms of the Laplace
transform of the waiting time distribution according to
\begin{eqnarray}
  \hat{\lambda}_{\mu_i} \left( u \right) & = & \frac{1}{u}  \frac{1 - \hat{f}
  (u)}{1 - \mu_i \hat{f} (u)} .  \label{eq:lu}
\end{eqnarray}
By linearity Eq.~(\ref{eq:elli}) determines the map, which is fixed by the
matrix elements
\begin{eqnarray}
  \rho_{11} \left( t \right) & = & \frac{1}{2} \left[ 1 + \lambda_z \left( t
  \right) \left( \rho_{11} \left( 0 \right) - \rho_{00} \left( 0 \right)
  \right) \right] \nonumber\\
  \rho_{10} \left( t \right) & = & \frac{1}{2} \left[ \rho_{10} \left( 0
  \right) \left( \lambda_x \left( t \right) - \lambda_y \left( t \right)
  \right) + \rho_{01} \left( 0 \right) \left( \lambda_x \left( t \right) +
  \lambda_y \left( t \right) \right) \right] . \nonumber
\end{eqnarray}
This provides the general solution of Eq.~(\ref{eq:ms}) for arbitrary $f
\left( t \right)$ and $\mathcal{E}$ a generic unital stochastic map. For later
use it is also convenient to express the map in a different superoperator
basis
\begin{eqnarray}
  \Lambda [\rho] & = & \sum^3_{k, l = 0} \Lambda'_{kl} \tau_k w \tau_l^{\dag},
  \label{eq:mappe2}
\end{eqnarray}
with $\Lambda'_{kl} = \sum_i \tmop{Tr}_S \{\tau_l \tau_i^{\dag} \tau_k \Lambda
[\tau_i]\}.$ This representation for superoperators was introduced in
{\cite{Sudarshan1961a}}, and apart from a multiplicative factor corresponding
to the space dimension it associates to the map its Choi matrix. For the
previous choice of basis set the associated matrix is still diagonal, so that
\begin{eqnarray}
  \Lambda \left( t, 0 \right) \left[ \rho \right] & = & \mu_0 \left( t \right)
  \rho + \mu_x \left( t \right) \sigma_x \rho \sigma_x + \mu_y \left( t
  \right) \sigma_y \rho \sigma_y + \mu_z \left( t \right) \sigma_z \rho
  \sigma_z,  \label{eq:mapp}
\end{eqnarray}
with $\tmmathbf{\mu} \left( t \right) = A^{- 1} \tmmathbf{\lambda} \left( t
\right) = \frac{1}{4} A\tmmathbf{\lambda} \left( t \right)$ and the $\left\{
\mu_i \left( t \right) \right\}$ at any time a probability distribution.

\subsection{Analysis of non-Markovianity}

We now study Markovianity of these quantum maps according to the criteria
introduced in {\cite{Breuer2009b,Rivas2010a}}. The first criterion is based on
the behavior of the trace distance among distinct initial states, which
quantifies how their distinguishability varies in time. For the considered map
the trace distance reads
\begin{eqnarray}
  D (\rho^1 (t), \rho^2 (t)) & = & \frac{1}{2} \| \rho^1 (t) - \rho^2 (t)\|_1
  \nonumber\\
  & = & \sqrt{\lambda^2_z \left( t \right) \Delta_p (0)^2 + \lambda_x^2
  \left( t \right) \Re \Delta_c (0)^2 + \lambda^2_y \left( t \right) \Im
  \Delta_c (0)^2}, \nonumber
\end{eqnarray}
where $\Delta_p (0)$ and $\Delta_c (0)$ denote population and coherence
differences at the initial time. The time derivative of this quantity, which
detects non-Markovianity identified with growth of trace distance in time for
at least a couple of possible initial states, is
\begin{eqnarray}
  \hspace{-2cm} \sigma (t, \rho^{1, 2} (0)) & = & \frac{\lambda_z \left( t
  \right) \dot{\lambda}_z \left( t \right) \Delta_p (0)^2 + \lambda_x \left( t
  \right) \dot{\lambda}_x^{} \left( t \right) \Re \Delta_c (0)^2 + \lambda_y
  \left( t \right) \dot{\lambda}_y \left( t \right) \Im \Delta_c
  (0)^2}{\sqrt{\lambda^2_z \left( t \right) \Delta_p (0)^2 + \lambda^2_x
  \left( t \right) \Re \Delta_c (0)^2 + \lambda^2_y \left( t \right) \Im
  \Delta_c (0)^2}}, \nonumber
\end{eqnarray}
so that a necessary and sufficient condition for non-Markovianity is that at
least one of the functions $\left| \lambda_i \left( t \right) \right|$ grows
in a certain time interval. This condition coincides with the requirement of
divisibility of the quantum map in terms of positive (but not necessarily
completely positive) maps according to Eq.~(\ref{eq:cpdiv}). For this time
evolution the map $\Lambda \left( t, s \right)$ is represented according to
Eq.~(\ref{eq:mape}) by the matrix
\begin{eqnarray}
  \Lambda_{kl} \left( t, s \right) & = & \tmop{diag} \left( \mathbbm{1},
  \frac{\lambda_x \left( t \right)}{\lambda_x \left( s \right)},
  \frac{\lambda_y \left( t \right)}{\lambda_y \left( s \right)},
  \frac{\lambda_z \left( t \right)}{\lambda_z \left( s \right)} \right),
  \nonumber
\end{eqnarray}
which corresponds to a positive map if and only if all eigenvalues lie between
zero and one. Let us now spell out these general relations for specific
choices of $\{ \mathcal{E}, f (t)\}$.

\begin{figure}[tb]
  \resizebox{60mm}{!}{\includegraphics{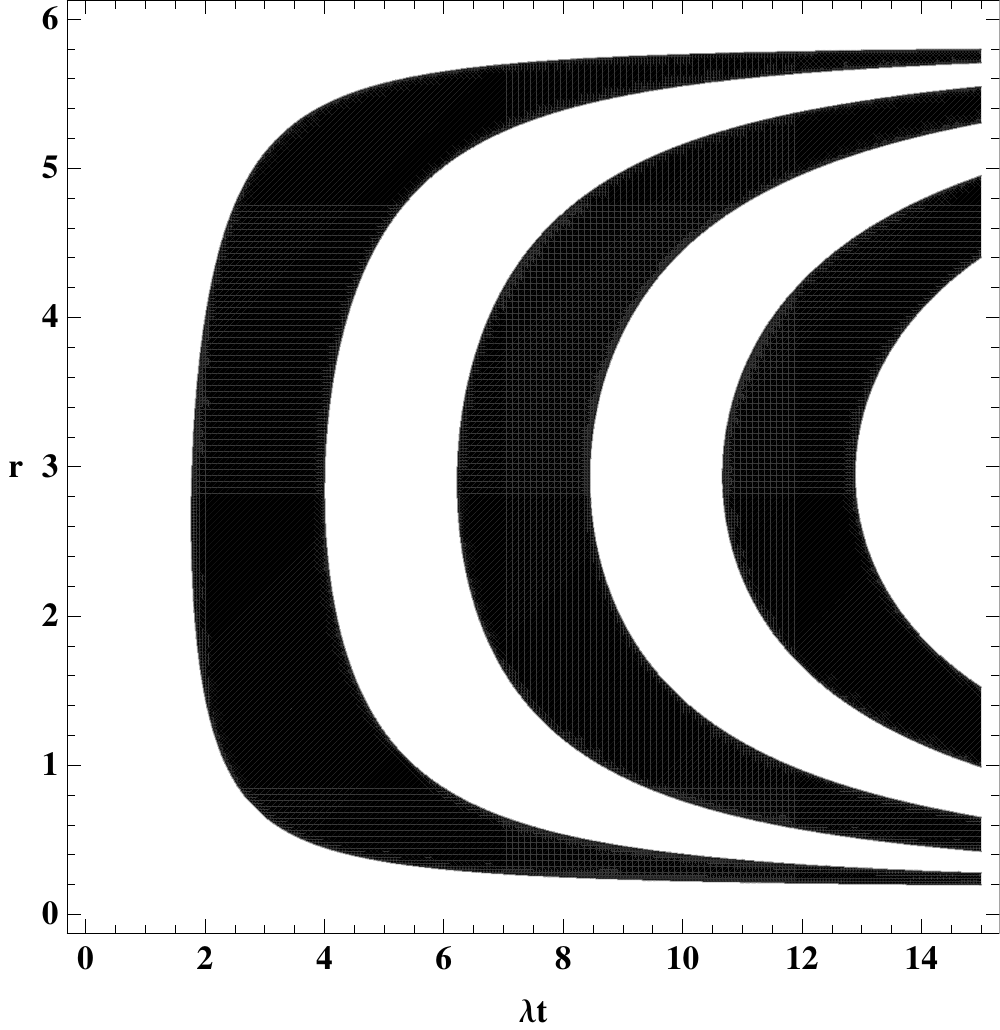}}{\hspace{1em}}\resizebox{60mm}{!}{\includegraphics{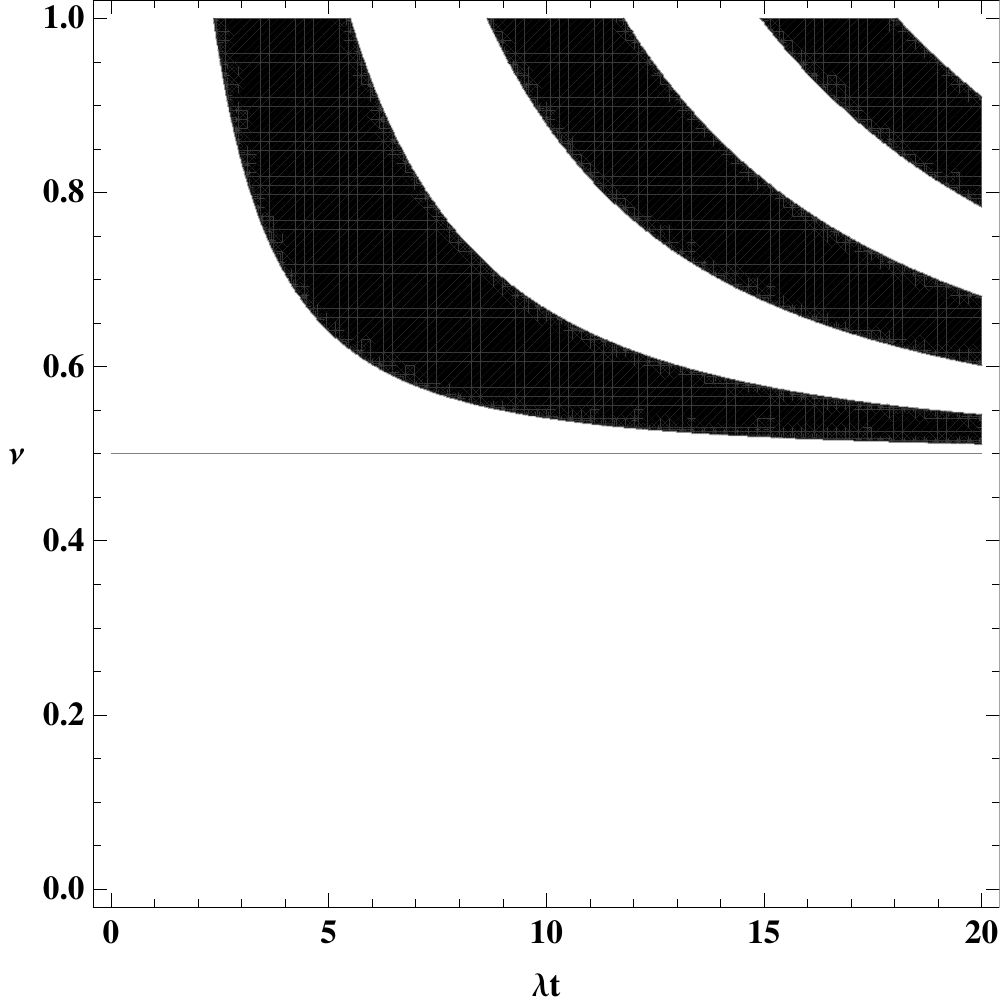}}
  \caption{\label{fig:contour}(Color online) (left) Plot of the sign of $q
  (t)$ given by Eq.~(\ref{eq:qged}) for \ the convolution of two exponential
  distributions with $\lambda_1 = \lambda$ and $\lambda_2 = r \lambda$ as a
  function of $r$ and $\lambda t$. The black regions correspond to a negative
  sign. (right) Plot of the sign of $\lambda_{1 - 2 \nu} (t)$ for the same
  waiting time distribution as a function of $\nu$ and $\lambda t$. One
  clearly sees the threshold for the inset of non-Markovianity at $\nu = 1 /
  2$.}
\end{figure}

For $\tmmathbf{\lambda}= \left( 0, 0, 0, 1 \right)$ we recover the phase-flip
channel
\begin{eqnarray}
  \mathcal{E}_z & = & \sigma_z \rho \sigma_z, \nonumber
\end{eqnarray}
which does not have a classical counterpart. For this case one has
$\tmmathbf{\mu}= \left( 1, - 1, - 1, 1 \right)$, leading to
$\tmmathbf{\lambda} \left( t \right) = \left( 1, q \left( t \right), q \left(
t \right), 1 \right)$, with $q \left( t \right)$ as in Eq.~(\ref{eq:qdiff}).
The map $\Lambda \left( t, 0 \right)$ describes pure dephasing, with
coherences multiplied by a factor $q \left( t \right)$. As a result the time
derivative of the quantifier of the distinguishability among the two evolved
states grows whenever $\left| q^{} (t) \right|$ grows in time. In such time
intervals, which we denote collectively by $\Omega_+ = \bigcup_i (a_i, b_i)$,
we have
\begin{eqnarray}
  \sigma (t, \rho^{1, 2} (0)) & \leqslant & | \Delta_c (0) |^{}
  \frac{\mathd}{\mathd t} |q (t) | \nonumber
\end{eqnarray}
so that the maximal growth is obtained for opposite equatorial states on the
Bloch sphere. The measure of non-Markovianity based on trace distance is then
given by
\begin{eqnarray}
  \mathcal{N} (\Lambda) & = & \int_{\Omega_+} \mathd t \frac{\mathd}{\mathd t}
  |q (t) | = \sum_i \left( \left| q \left( b_i \right) \right| - \left| q
  \left( b_i \right) \right| \right) . \nonumber
\end{eqnarray}
To consider an interesting class of situations we now specify also the waiting
time distribution, considering the convolution of $m$ exponential waiting time
distributions leading to so called Erlang distributions of the form $f_m
\left( t \right) = \lambda \frac{(\lambda t)^{m - 1}}{(m - 1) !} \mathe^{-
\lambda t}$, for which the difference of the probabilities to have an even and
an odd number of jumps becomes
\begin{eqnarray}
  q_m (t) & = & \mathe^{- \lambda t} \sum_{n = 0}^{\infty} \left( - \right)^n
  \sum_{k = 0}^{m - 1} \frac{(\lambda t)^{mn + k}}{\left( mn + k \right) !} .
  \nonumber
\end{eqnarray}
For the memoryless case $m = 1$ one has the strictly monotone decreasing
function $q_1 (t) = \exp \left( - 2 \lambda t \right)$ according to
Eq.~(\ref{eq:gfp}), while for $m = 2$ one obtains
\begin{eqnarray}
  q_2 (t) & = & \mathe^{- \lambda t} [\cos (\lambda t) + \sin (\lambda t)],
  \nonumber
\end{eqnarray}
which oscillates and crosses zero at isolated points. For $m \geqslant 2$
these functions exhibit an oscillating behavior, so that the minima of $|q_m
(t) |$ lie on the real axis. The modulus of these functions for values of $m$
up to six is shown in Fig.~\ref{fig:kolmogorov-qm}. For each choice of waiting
time distribution $f_m (t)$ the measure of non-Markovianity is given by the
series
\begin{eqnarray}
  \mathcal{N} \left( \Lambda_m \right) & = & \sum_{t_i \in M} \left| q_m
  \left( t_i \right) \right|,  \label{eq:mesqm}
\end{eqnarray}
where $M$ is the denumerable set of times corresponding to the maxima of the
function $|q_m |$. An exact analytic evaluation of the measure is feasible for
$m = 2$, since the maxima correspond to $\lambda t = n \pi$, with $n$ a
positive integer, so that
\begin{eqnarray}
  \mathcal{N} \left( \Lambda_2 \right) & = & \sum^{\infty}_{n = 1} \mathe^{- n
  \pi} = \frac{1}{e^{\pi} - 1} . \nonumber
\end{eqnarray}
More generally as shown in Fig.~\ref{fig:kolmogorov-qm} the first maximum of
$|q_m |$ is above the first maximum of $|q_n |$ for $m > n$, and the same
occurs for the other maxima, whose values decrease exponentially. This
substantiates the statement that waiting time distributions corresponding to
the convolution of a higher number of exponential terms have stronger memory
{\cite{Cox1965}}. Indeed in such a case the overall waiting time is the sum of
a number $m$ of waiting times, which can be thought in series. While each of
them is described by a memoryless probability distribution, the overall
distribution deviates from the memoryless case, the more so the higher the
number of terms. Another situation in which Markovianity or non-Markovianity
can be observed is given considering the convolution of different exponential
waiting time distributions. In this case we have $\mathsf{f} = f_1 \ast f_2$,
where each $f_i$ is of the form Eq.~(\ref{eq:expwtd}) with parameter
$\lambda_i$, which leads to
\begin{eqnarray}
  \mathsf{f} (t) & = & 2 \frac{p}{s} \mathe^{- \frac{1}{2} st}
  \frac{1}{\sqrt{1 - 4 \frac{p}{s^2}}} \sinh ( \frac{st}{2} \sqrt{1 - 4
  \frac{p}{s^2}})  \label{eq:fged}
\end{eqnarray}
where $s$, $p$ denote sum and product of the parameters $\lambda_i$. Setting
$\lambda_1 = \lambda$ and $\lambda_2 = r \lambda$, so that the parameter $r =
\lambda_2 / \lambda_1$ gives the relative scale among the rates of the two
waiting time distributions the function $q (t)$ takes the
\begin{eqnarray}
  \hspace{-2.5cm} q \left( t \right) & = & \mathe^{- \frac{1 + r}{2} \lambda
  t} \left[ \cosh \left( \sqrt{r^2 - 6 r + 1} \frac{\lambda t}{2} \right) +
  \frac{1 + r}{\sqrt{r^2 - 6 r + 1}} \sinh \left( \sqrt{r^2 - 6 r + 1}
  \frac{\lambda t}{2} \right) \right],  \label{eq:qged}
\end{eqnarray}
which can oscillate and take on negative values, thus leading to a non zero
measure of non-Markovianity, for $3 - 2 \sqrt{2} \leqslant r \leqslant 1 / (3
- 2 \sqrt{2})$. Also in this case non-Markovianity does not appear when one of
the rates is much stronger than the other, so that the overall distribution is
dominated by only one of the two components, which is memoryless distributed.
The sign of $q (t)$ as a function of the relative strength $r$ and the
rescaled time $\lambda t$ is plotted in
Fig.~\ref{fig:contour}. Note that also in this case the measure of non-Markovianity is
given by the sum of the maxima of $|q (t) |$.

The value $\tmmathbf{\lambda}= \left( 0, \frac{1}{2}, \frac{1}{2}, 0 \right)$
leads to the map
\begin{eqnarray}
  \mathcal{E}_{_P} \rho & = & \frac{1}{2} \sigma_x \rho \sigma_x + \frac{1}{2}
  \sigma_y \rho \sigma_y = \sigma_+ \rho \sigma_- + \sigma_- \rho \sigma_+,
  \nonumber
\end{eqnarray}
for which also populations are affected, since $\tmmathbf{\mu}= \left( 1, 0,
0, - 1 \right)$ and $\tmmathbf{\lambda}= \left( 1, g \left( t \right), g
\left( t \right), q \left( t \right) \right)$, with $g \left( t \right)$
survival probability as in Eq.~(\ref{eq:k-sur}). As discussed in
{\cite{Vacchini2011a}} the measure of non-Markovianity of this map according
to the trace distance criterion is the same as for the dephasing map
characterized by $\tmmathbf{\lambda}= \left( 0, 0, 0, 1 \right)$, so that in
both cases it is determined by the waiting time distribution. As we shall see
this is no more true for the other measure of quantum non-Markovianity.

In the case $\tmmathbf{\lambda}= \left( 1 - \nu, 0, 0, \nu \right)$, with $0
\leqslant \nu \leqslant 1$, one has a mixture of phase-flip and identity,
leading to $\tmmathbf{\mu}= \left( 1, 1 - 2 \nu, 1 - 2 \nu, 1 \right)$ and
therefore $\tmmathbf{\lambda} \left( t \right) = \left( 1, \lambda_{1 - 2 \nu}
\left( t \right), \lambda_{1 - 2 \nu} \left( t \right), 1 \right)$, with
$\hat{\lambda}_{1 - 2 \nu} \left( u \right)$ as in Eq.~(\ref{eq:lu}). The
resulting map again describes pure dephasing, but the off-diagonal matrix
element is multiplied by a factor $\lambda_{1 - 2 \nu} \left( t \right)$. In
order to assess non-Markovianity one now has to study $\lambda_{1 - 2 \nu}
\left( t \right)$ instead of $q \left( t \right)$. Considering again the same
class of waiting time distributions, one sees that now a threshold appears, so
that one can have non-Markovianity only if $\nu > \frac{1}{2}$, as shown in
Fig.~\ref{fig:contour}. This shows that indeed non-Markovianity depends on
both elements of the couple $\left\{ \mathcal{E}, f \left( t \right)
\right\}$, at variance with the classical case and the previous examples.

\begin{figure}[tb]
  \resizebox{!}{60mm}{\includegraphics{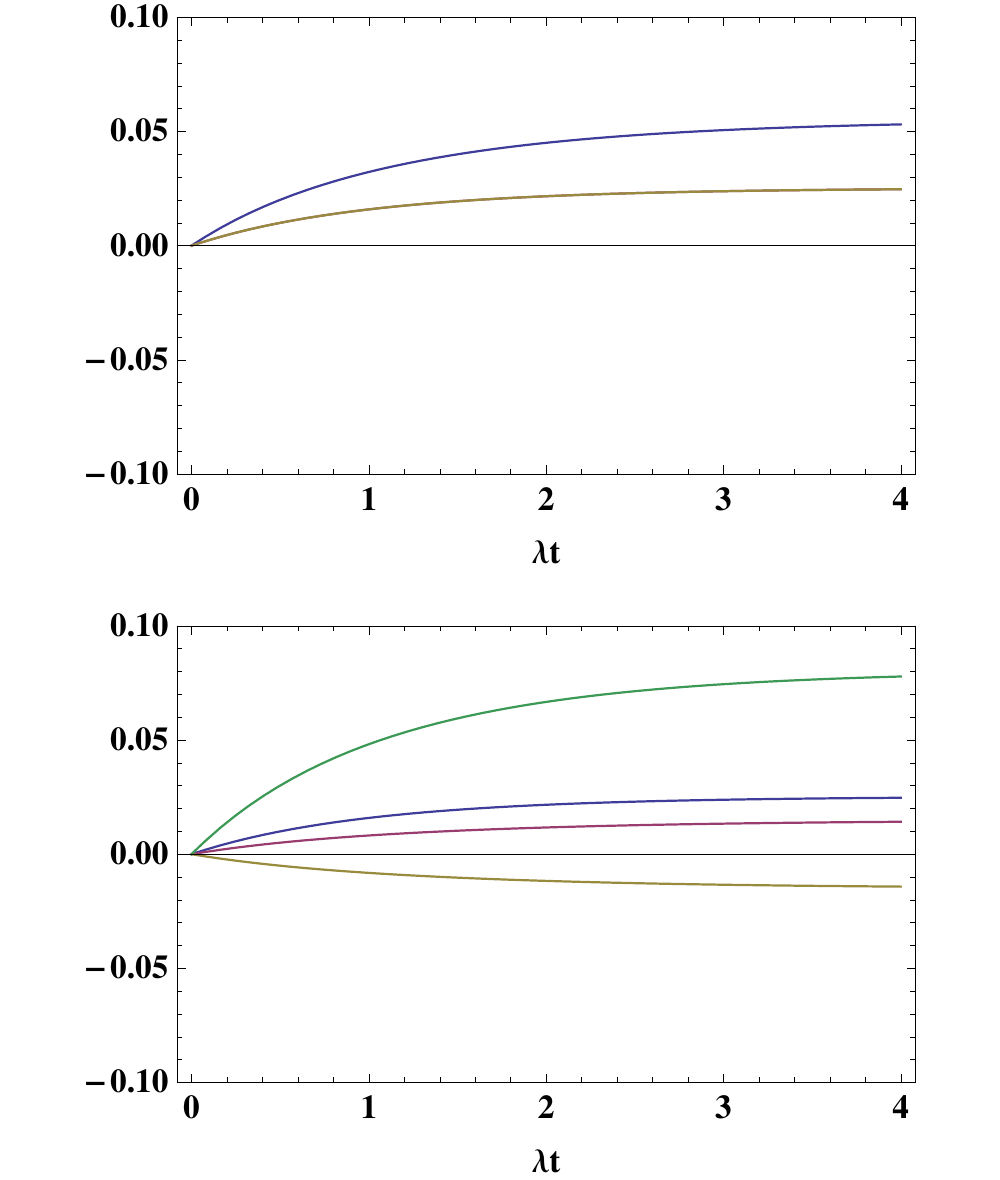}}{\hspace{1em}}\resizebox{60mm}{!}{\includegraphics{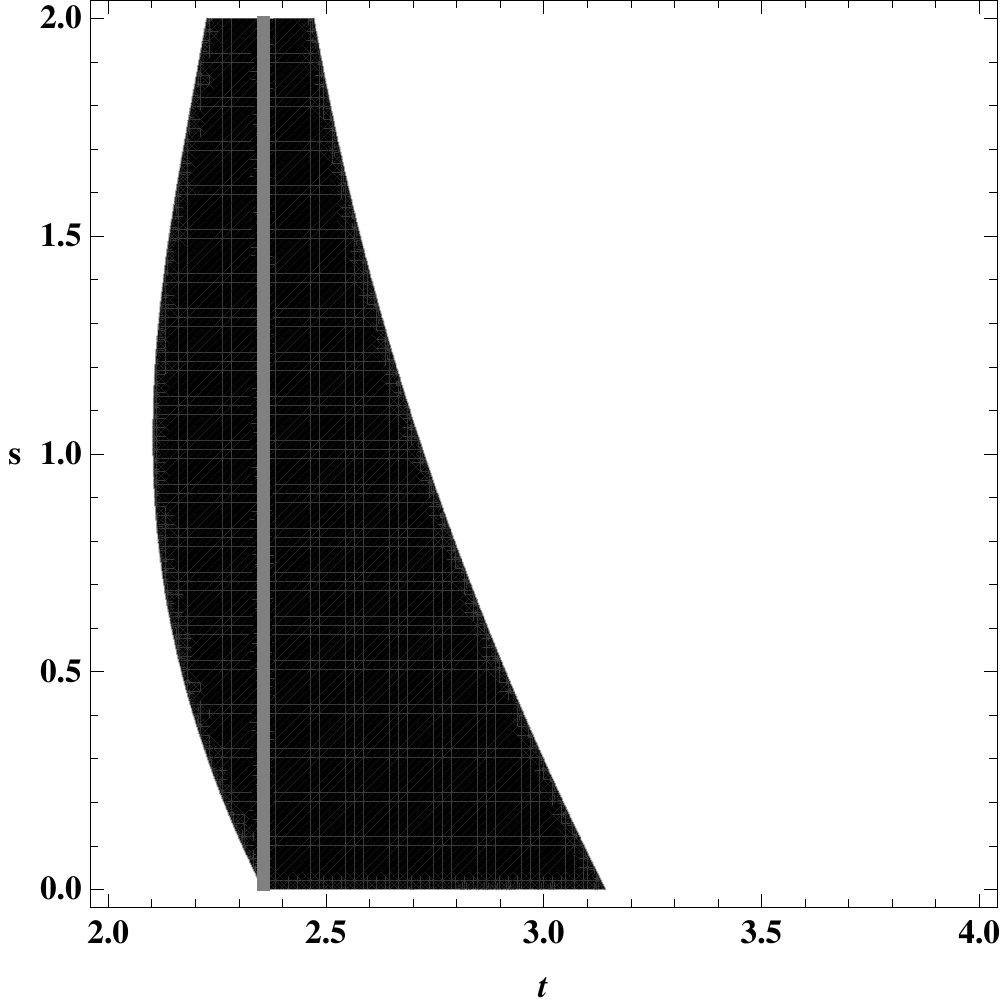}}
  \caption{\label{fig:channels-choi}(Color online) (top left) Plot of the two
  distinct coefficients of Eq.~(\ref{eq:ok}). (bottom left) Plot of the four
  distinct coefficients of Eq.~(\ref{eq:wrong}). In the latter case one is
  always negative. The coefficients of the two equivalent expressions of the
  time-convolutionless master equation associated to the map
  Eq.~(\ref{eq:mapp}) for the case $\tmmathbf{\lambda}= \left( \frac{1}{5},
  \frac{2}{5}, \frac{1}{5}, \frac{1}{5} \right)$, so that $\tmmathbf{\mu}=
  \left( 1, \frac{1}{5}, - \frac{1}{5}, - \frac{1}{5} \right)$ are obtained
  from Eq.~(\ref{eq:lt}). The considered waiting time distribution is of the
  form Eq.~(\ref{eq:fged}) with $\lambda_2 = r \lambda$ and $r = 0.13$.
  (right) Plot of the sign of the lowest non zero eigenvalue of the Choi
  matrix corresponding to the map $\Lambda (t + s, t)$, for a stochastic map
  corresponding to pure dephasing and waiting given by the convolution of two
  equal exponentials. The black regions correspond to a negative sign. For
  certain values of $t$, corresponding to the points in which the inverse does
  not exist, in the plot denoted by a gray vertical line, the memory is
  arbitrary long, while it stays finite for other values.}
\end{figure}

The criterion of non-Markovianity introduced in {\cite{Rivas2010a}} relies on
the requirement of divisibility of the quantum map in terms of completely
positive maps, so that Eq.~(\ref{eq:cpdiv}) applies where now each $\Lambda$
is a completely positive map. To detect the failure of divisibility one can
consider the Choi matrix associated to the map $\Lambda \left( t, s \right)$,
with $t \geqslant s$, which can be obtained from Eq.~(\ref{eq:mappe2}). For
the general case the expression is somewhat cumbersome, and an alternative way
is to look at the sign of the coefficients in the master equation in
time-convolutionless form corresponding to Eq.~(\ref{eq:ms}). For the case
$\tmmathbf{\lambda}= \left( 0, \frac{1}{2}, \frac{1}{2}, 0 \right)$ this path
has been followed in {\cite{Vacchini2011a}}, the master equation reads
\begin{eqnarray}
  \hspace{-2cm} \frac{d \rho}{\tmop{dt}} & = & - \frac{1}{4} \left(
  \frac{\dot{\lambda}_x \left( t \right)}{\lambda_x \left( t \right)} +
  \frac{\dot{\lambda}_y \left( t \right)}{\lambda_y \left( t \right)} -
  \frac{\dot{\lambda}_z \left( t \right)}{\lambda_z \left( t \right)} \right)
  \left[ \sigma_z \rho \sigma_z - \rho \right] - \frac{1}{2}
  \frac{\dot{\lambda}_z \left( t \right)}{\lambda_z \left( t \right)} \left[
  \sigma_+ \rho \sigma_- + \sigma_- \rho \sigma_+ - \rho \right] \nonumber
\end{eqnarray}
with $\lambda_z \left( t \right) = q \left( t \right)$, $\lambda_x \left( t
\right) = \lambda_y \left( t \right) = g \left( t \right)$. In this setting
one can consider situations in which the measure of non-Markovianity related
to the trace distance is zero, since the map is divisible in the sense of
positive maps, but the intermediate maps are not completely positive. This is
the case for the convolution of two different exponential waiting time
distributions as in Eq.~(\ref{eq:fged}), for a suitable choice of the ratio
$\lambda_1 / \lambda_2$. Note that for the situation in which all $\lambda_i
\left( t \right)$ are different one has
\begin{eqnarray}
  \hspace{-2.5cm} \frac{d \rho}{\tmop{dt}} & = & - \frac{1}{4} \left(
  \frac{\dot{\lambda}_x \left( t \right)}{\lambda_x \left( t \right)} +
  \frac{\dot{\lambda}_y \left( t \right)}{\lambda_y \left( t \right)} -
  \frac{\dot{\lambda}_z \left( t \right)}{\lambda_z \left( t \right)} \right)
  \left[ \sigma_z \rho \sigma_z - \rho \right] - \frac{1}{2}
  \frac{\dot{\lambda}_z \left( t \right)}{\lambda_z \left( t \right)} \left[
  \sigma_+ \rho \sigma_- + \sigma_- \rho \sigma_+ - \rho \right] 
  \label{eq:wrong}\\
  \hspace{-2.5cm} &  & + \frac{1}{4} \left( \frac{\dot{\lambda}_x \left( t
  \right)}{\lambda_x \left( t \right)} - \frac{\dot{\lambda}_y \left( t
  \right)}{\lambda_y \left( t \right)} \right) \left[ \sigma_x \rho \sigma_x -
  \rho \right] - \frac{1}{4} \left( \frac{\dot{\lambda}_x \left( t
  \right)}{\lambda_x \left( t \right)} - \frac{\dot{\lambda}_y \left( t
  \right)}{\lambda_y \left( t \right)} \right) \left[ \sigma_y \rho \sigma_y -
  \rho \right], \nonumber
\end{eqnarray}
and the two intermediate channels have opposite coefficients, so that unless
$\lambda_x \left( t \right) = \lambda_y \left( t \right)$, which is the case
considered in {\cite{Vacchini2011a}}, one is always negative. In this case
however one cannot read divisibility from the sign of the coefficients, since
the Lindblad operators appearing in them are not linearly independent, indeed
Eq.~(\ref{eq:wrong}) can be written as
\begin{eqnarray}
  \hspace{-2.5cm} \frac{d \rho}{\tmop{dt}} & = & + \frac{1}{4} \left(
  \frac{\dot{\lambda}_x \left( t \right)}{\lambda_x \left( t \right)} -
  \frac{\dot{\lambda}_y \left( t \right)}{\lambda_y \left( t \right)} -
  \frac{\dot{\lambda}_z \left( t \right)}{\lambda_z \left( t \right)} \right)
  \left[ \sigma_x \rho \sigma_x - \rho \right]  \label{eq:ok}\\
  \hspace{-4cm} &  & \hspace{-1cm} - \frac{1}{4} \left( \frac{\dot{\lambda}_x
  \left( t \right)}{\lambda_x \left( t \right)} - \frac{\dot{\lambda}_y \left(
  t \right)}{\lambda_y \left( t \right)} + \frac{\dot{\lambda}_z \left( t
  \right)}{\lambda_z \left( t \right)} \right) \left[ \sigma_y \rho \sigma_y -
  \rho \right] - \frac{1}{4} \left( \frac{\dot{\lambda}_x \left( t
  \right)}{\lambda_x \left( t \right)} + \frac{\dot{\lambda}_y \left( t
  \right)}{\lambda_y \left( t \right)} - \frac{\dot{\lambda}_z \left( t
  \right)}{\lambda_z \left( t \right)} \right) \left[ \sigma_z \rho \sigma_z -
  \rho \right] . \nonumber
\end{eqnarray}
In the present framework we can indeed point to a situation in which, due to a
subtle balance, all coefficients in Eq.~(\ref{eq:ok}) are positive, thus
granting divisibility, but this is no more true for the coefficients of
Eq.~(\ref{eq:wrong}). This situation is depicted in
Fig.~\ref{fig:channels-choi} considering $\tmmathbf{\lambda}= \left( \frac{1}{5}, \frac{2}{5},
\frac{1}{5}, \frac{1}{5} \right)$ and a suitable waiting time distribution.
Even though this fact can be easily understood from a conceptual point of
view, it is useful to stress it by means of an explicit example.

For the case $\tmmathbf{\lambda}= \left( 0, 0, 0, 1 \right)$ of pure dephasing
in order to detect non-Markovianity it is convenient to consider the Choi
matrices associated to the maps $\Lambda (t + s, t) = \Lambda (t + s, 0)
\Lambda^{- 1} (t, 0)$, so that according to Eq.~(\ref{eq:mappe2}) we obtain
\begin{eqnarray}
  \Lambda'_{kl} (t + s, t) & = & \tmop{diag} \left( 1 + \frac{q (t + s)}{q
  (t)}, 0, 0, 1 - \frac{q (t + s)}{q (t)} \right),  \label{eq:diag}
\end{eqnarray}
where the simplicity of the result strongly depends on the convenient choice
of basis. Failure of divisibility is then detected when at least one of the
coefficients of these collection of matrices depending on two temporal indexes
becomes negative. The sign of the smallest non zero eigenvalue is plotted in
Fig.~\ref{fig:channels-choi} for a waiting time given by the convolution of
two identical exponential distributions. In accordance to the result obtained
relying on the criterion based on trace distance distinguishability, the map
is indeed non-Markovian. Note however that the divisibility property does
depend on the initial time considered, i.e. for certain time windows, which
obviously include the initial time $t = 0$, the maps $\Lambda (t + s, t)$ are
completely positive for any time interval $s$. At the same time the violation
of complete positivity does decrease for large $s$. Moreover while the two
criteria agree in labelling the map as non-Markovian, as discussed in detail
in {\cite{Vacchini2011a}} they assign to it different measures. In particular
as follows from Eq.~(\ref{eq:diag}) the approach based on divisibility assigns
an infinite measure to this map as soon as the quantity $q (t)$ goes through
zero, so that at variance with Eq.~(\ref{eq:mesqm}) processes with different
memories are put on the same footing. Let us note that this difficulty can be
overcome by keeping divisibility of the time evolution in terms of completely
positive maps as a signature of Markovianity, quantifying however its
violation in a different way. As suggested in {\cite{Hou2011a}} one can
introduce a different weight, considering the integral of the arcotangent of
the sum of the negative eigenvalues of the matrix Eq.~(\ref{eq:diag}) in the
time regions in which complete positivity breaks down, renormalizing by the
extension of these regions. For the case at hand indeed this modification
makes the non-Markovianity measure based on divisibility finite, even though
the expression Eq.~(\ref{eq:mesqm}) for the measure based on
distinguishability remains much easier to evaluate.

\section{Conclusions\label{sec:ceo}}

In the present manuscript we have considered a class of quantum dynamics which
can be obtained from a generic unital stochastic map on $\mathbbm{C}^2$ and a
classical waiting time distribution, thus extending the class of examples
considered in Ref.~{\cite{Vacchini2011a}} to show how the recently introduced
notions of quantum non-Markovianity relate to the classical one. The
considered examples show how versatile the class of semi-Markov processes and
their quantum counterpart can be in order to study the notion of non-Markovian
process in the quantum framework and highlight different possible behavior.

\section*{Acknowledgments}

The author thanks A. Smirne for discussions and reading of the manuscript.
This work was supported by MIUR under PRIN 2008, and COST under MP1006.

\end{document}